\documentclass[3p,times,procedia]{elsarticle}
\usepackage{ecrc}
\usepackage{algorithm}
\usepackage{algorithmic}
\usepackage{subfigure}
\usepackage{graphicx}
\usepackage[section] {placeins}
\usepackage{multirow}
\usepackage{amsmath}
\usepackage{amssymb}
\volume{00}
\firstpage{1}
\journalname{Procedia Computer Science}
\runauth{}
\jid{procs}
\jnltitlelogo{Procedia Computer Science}
\CopyrightLine{2013}{Published by Elsevier Ltd.}
\usepackage{amssymb}
\usepackage[figuresright]{rotating}
\begin{document}
\begin{frontmatter}
\dochead{International Workshop on Body Area Sensor Networks (BASNet-2013)}
\title{BEENISH: Balanced Energy Efficient Network Integrated Super Heterogenous Protocol for Wireless Sensor Networks}
\author{T. N. Qureshi$^{\pounds}$, N. Javaid$^{\pounds}$, A. H. Khan$^{\pounds}$, A. Iqbal$^{\pounds}$, E. Akhtar$^{\sharp}$, M. Ishfaq$^{\S}$}
\address{$^{\pounds}$COMSATS Institute of Information Technology, Islamabad, Pakistan. \\
        $^{\sharp}$University of Bedfordshire, Luton, UK.\\
        $^{\S}$King Abdulaziz University, Rabigh, Saudi Arabia.}

\begin{abstract}
In past years there has been increasing interest in field of Wireless Sensor Networks (WSNs). One of the major issue of WSNs is development of energy efficient routing protocols. Clustering is an effective way to increase energy efficiency. Mostly, heterogenous protocols consider two or three energy level of nodes. In reality, heterogonous WSNs contain large range of energy levels. By analyzing communication energy consumption of the clusters and large range of energy levels in heterogenous WSN, we propose BEENISH (Balanced Energy Efficient Network Integrated Super Heterogenous) Protocol. It assumes WSN containing four energy levels of nodes. Here, Cluster Heads (CHs) are elected on the bases of residual energy level of nodes. Simulation results show that it performs better than existing clustering protocols in heterogeneous WSNs. Our protocol achieve longer stability, lifetime and more effective messages than Distributed Energy Efficient Clustering (DEEC), Developed DEEC (DDEEC) and Enhanced DEEC (EDEEC).
\end{abstract}
\end{frontmatter}

\section{Introduction}
Wireless Sensor Networks (WSNs) \cite{Reference 1,Reference 18,Reference 19} have become popular in variety of applications such as military surveillance, environmental, transportation traffic, temperature, pressure and vibration monitoring. To achieve fault tolerance, WSNs consist of hundreds or even thousands of sensor randomly distributed with in the region \cite{Reference 2,Reference 20,Reference 21}. All the nodes report sensed data to Base Station (BS) often called sink. Nodes in WSNs are power constrained due to limited battery resource, and they might be placed where they can not be accessed, so,impossible to recharge or replace. To save energy, regular and long distance communication should be avoided to prolong network lifetime \cite{Reference 1}. Sensor nodes take self decisions to accomplish sensing tasks, constructing network topology and routing policy. Therefore, it become important to design energy efficient algorithm for enhancing robustness against node failures and extending lifetime of WSNs.

Efficiently Grouping sensor nodes in form of clusters is beneficial in minimizing the energy utilization. Numerous energy efficient protocols are made based on clustering structure\cite{Reference 1,Reference 5,Reference 6}. In clustering, nodes assemble themselves in form of clusters with one node acting as the Cluster Head (CH). All cluster member nodes transmit sensed data to their CH, while the CH aggregate data received and forward it to the remote BS \cite{Reference 7,Reference 8}. Clustering can be formed in two kind of networks i.e., homogenous and heterogeneous. WSNs having nodes of same energy level are called homogenous WSNs. Low Energy Adaptive Clustering Hierarchy (LEACH) \cite{Reference 9}, Power Efficient Gathering in Sensor Information Systems (PEGASIS) \cite{Reference 10} and Hybrid Energy-Efficient Distributed Clustering (HEED) \cite{Reference 11} are examples of cluster based protocols which are designed for homogenous WSNs. These algorithms poorly perform in heterogeneous regions. Nodes have less energy will expire faster than high energy nodes because these homogenous clustering based algorithms are incapable to treat every node with respect to energy. In heterogeneous WSNs, nodes are deployed with different initial energy levels. Heterogeneity in WSN may be the result of re-energizing of WSN in order to extend the network lifetime \cite{Reference 12,Reference 16,Reference 17}. Stable Election Protocol (SEP) \cite{Reference 12}, Distributed Energy Efficient Clustering (DEEC) \cite{Reference 13}, Developed DEEC (DDEEC) \cite{Reference 14}, Enhanced DEEC (EDEEC) \cite{Reference 15} are protocols for heterogenous WSNs.

\section{Radio Dissipation Model}
The radio energy model describes that l-bit message is transmitted over a distance $d$ as in \cite{Reference 8,Reference 9} as shown in Fig. 1.

\begin{figure}[!h]
\centering
\includegraphics[height=2cm, width=11cm]{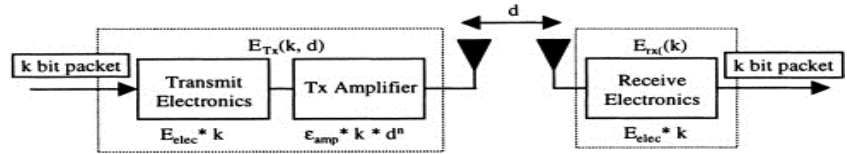}
\caption{Radio Energy Dissipation Model}
\end{figure}

\begin{eqnarray}
E_{Tx}(l,d)=
\begin{cases}
l E_{elec}+l\varepsilon_{fs}d^{2},&  d<d_{0} \\
l E_{elec}+l \varepsilon_{mp}d^{4},& d\geq d_{0} \\
\end{cases}
\end{eqnarray}
\\

Where $E_{elec}$ is energy used per bit to run transmitter or receiver circuit. Free space $(fs)$ model is used if distance is in less than threshold otherwise multi path $(mp)$ model.
Now, total energy dissipated in the network during a round is given below, as supossed \cite{Reference 8,Reference 9}.
\begin{eqnarray}
E_{round}= L(2NE_{elec}+NE_{DA}+k\varepsilon_{mp}d_{to BS}^{4}+N\varepsilon_{fs}d_{to CH}^{2})
\end{eqnarray}
\\

Where, $k$= number of clusters,\\
$E_{DA}$= Data aggregation cost expended in CH\\
$d_{toBS}$= Average distance between CH and BS\\
$d_{toCH}$= Average distance between cluster members and CH\\

Assuming all nodes are uniformly distributed over network so, $d_{to BS}$ and $d_{to CH}$ can be calculated as following as in  \cite{Reference 8,Reference 9}:
\begin{eqnarray}
d_{to CH}= \frac{M}{\sqrt{2 \pi k}}, d_{to BS}= 0.765\frac{M}{2}
\end{eqnarray}

By finding the derivative of $E_{Round}$ with respect to $k$ to zero, we get the $k_{opt}$ optimal number clusters as in \cite{Reference 8,Reference 9,Reference 13}.

\begin{eqnarray}
k_{opt}=\frac{\sqrt{N}}{\sqrt{2\pi}}\sqrt{\frac{\varepsilon_{fs}}{\varepsilon_{mp}}}\frac{M}{d_{toBS}^{2}}
\end{eqnarray}

\section{The BEENISH Protocol}
In this section, we present details of our BEENISH protocol. BEENISH implements the same concept as in DEEC, in terms of selecting CH which is based on residual energy level of the nodes with respect to average energy of network. However, DEEC is based on two types of nodes; normal and advance nodes. BEENISH uses the concept of four types of nodes; normal, advance, super and ultra-super nodes.

Let $n_{i}$ shows the rounds for a node $s_{i}$ to become CH, we refer it as rotating epoch. CH has to consume more energy as compare to member nodes. In homogeneous networks, to ensure average $p_{opt} N$ CHs in each round, LEACH let every node $s_{i} (i=1,2,....N)$ to become CH once in every $n_{i}=\frac{1}{p_{opt}}$ rounds. During operation of WSN all the nodes does not own the same remaining energy. So, if the epoch  $n_{i}$ is kept equal for all nodes as in LEACH then energy is not efficiently distributed and nodes having low energy die before high energy nodes. BEENISH choose different epoch $n_{i}$ for different nodes with respect to their remaining energy $E_{i} (r)$. High energy nodes are more often elected as CH as compare to low energy nodes. So, high energy nodes have smaller epoch $n_{i}$ as compare to high energy nodes . In BEENISH ultra-super nodes are largely elected as CH as compare to super, advance and normal nodes, and so, on. In this way energy consumed by all nodes is equally distributed.

Let $p_{i}=\frac{1}{n_{i}}$ is probability of node to become CH during epoch $n_{i}$ rounds. When all the nodes have same every level at each epoch, selecting the average probability $p_{i}$ to be $p_{opt}$ can ensure that there are $p_{opt} N$ CHs every round and approximately all nodes die at the same time. If nodes are having different energy then nodes with more energy have $p_{i}$ larger than $p_{opt}$.

In BEENISH, average energy of $r^{th}$ round can be obtained as follows and as supposed in DEEC:

\begin{eqnarray}
\bar{E}(r)= \frac{1}{N}E_{total}(1-\frac{r}{R})
\end{eqnarray}

$R$ is showing total rounds from the start of network to the all nodes die and can be estimated as in DEEC and given as under:

\begin{eqnarray}
R= \frac{E_{total}}{E_{round}}
\end{eqnarray}

$E_{round}$ is the energy dissipated in a network during single round as given in 2.


%
%
%
%

To achieve the optimal number of CH at start of each round, node $s_{i}$ decides whether to become a CH or not based on probability threshold calculated by expression in the following equation, and as supposed in  \cite{Reference 9,Reference 13}.

\begin{eqnarray}
T(s_{i})=
\begin{cases}
\frac{p_{i}}{1-p_{i}(rmod\frac{1}{P_{i}})} & if\; s_{i}\;\epsilon \;G \\
0 & otherwise
\end{cases}
\end{eqnarray}

Where $G$ is the set of nodes eligible to become CH. If a node $s_{i}$ has not been CH in the most recent $n_{i}$ then it belongs to set $G$. Random number between 0 and 1 is selected by nodes belonging to set $G$. If the number is less than threshold $T(s_{i})$, the node $s_{i}$ will be CH for that current round.

In real scenarios, WSNs have more greater than two or three energy levels of nodes. In WSN due to random CH selection, large range of energy levels are created. So, as much more energy levels we quantize and define different probability for every energy level will lead to as much better results and lead to energy efficiency. In BEENISH, we first time use concept of four level heterogeneous network having normal, advance, super and ultra-super nodes. The probabilities for four types of nodes are given below:

\begin{eqnarray}
p_{i}=
\begin{cases}
\frac{p_{opt}E_{i}(r)}{(1+m(a+m_{0}(-a+b+m_{1}(-b+u))))\bar{E}(r)}   & s_{i} \;is\; the\; normal \;node\\
\frac{p_{opt}(1+a)E_{i}(r)}{(1+m(a+m_{0}(-a+b+m_{1}(-b+u))))\bar{E}(r)}  & s_{i}\; is \;the \;advanced \;node\\
\frac{p_{opt}(1+b)E_{i}(r)}{(1+m(a+m_{0}(-a+b+m_{1}(-b+u))))\bar{E}(r)}  & s_{i} \; is \;the\; super \;node\\
\frac{p_{opt}(1+u)E_{i}(r)}{(1+m(a+m_{0}(-a+b+m_{1}(-b+u))))\bar{E}(r)}  & s_{i} \; is \;the\; ultra-super \;node\\
\end{cases}
\end{eqnarray}

Threshold is calculated for CH selection of normal, advanced, super and ultra-super nodes by putting above values in equation below.

\begin{eqnarray}
T(s_{i})=
\begin{cases}
\frac{p_{i}}{1-p_{i}(rmod\frac{1}{P_{i}})} & if\; s_{i}\;\epsilon \;G \\
0 & otherwise
\end{cases}
\end{eqnarray}

In the equation of $T(s_{i})$, we find that nodes with greater remaining energy $E_{i}(r)$ at round $r$ are more possibly to become CH as compare to low energy nodes. The aim of this mechanism is to efficiently divide the energy consumption in the network and extend the stability period which is defined by first node die and network lifetime defined by last node die from the start of WSN.

Simulations show that BEENISH is more efficient protocol than DEEC, DDEEC and EDEEC for WSN containing four and multi level heterogeneity in terms of first node die and last node die.

\section{Simulations and Results}
This section evaluates the performance of BEENISH protocol using MATLAB. We consider a WSN containing of $N=100$ nodes randomly deployed inside $100m\times100m$ field. For simplicity, we assume that all nodes are either fixed or micro-mobile and ignore energy loss due to signal collision and interference between signals of different nodes that are due to dynamic random channel conditions. Our simulations use radio parameters mentioned in Table 1. Protocols compared with BEENISH include DEEC, DDEEC and EDEEC. We estimated performance for the case of four level and multi-level heterogeneous WSNs. We observe performance of DEEC, DDEEC, EDEEC and BEENISH for four level and multi-level heterogenous WSNs. We take the parameters; $m=0.5, \;m_{0}=0.3, \;m_{1}=0.2, \;a=1.5, \;b=2.0 \;and \;u=2.5$, containing $50$ normal nodes having $E_{0}$ energy, $35$ advanced nodes having $1.5$ times more energy than normal nodes, $12$ super nodes containing $2$ times more energy than normal nodes and $3$ ultra-super nodes containing $2.5$ times more energy than normal nodes. First node for DEEC, DDEEC, EDEEC and BEENISH dies at 1103, 1367, 1421 and 1661 rounds, respectively. All nodes die at 5191, 3976, 6866 and 6903 rounds, respectively. Fig. 3 shows BEENISH sends more data to BS than DEEC, DDEEC and EDEEC. BEENISH is efficient as compare to all protocols in terms of stability period, network life time and packets sent to the BS.

\begin{figure}[!h]
\centering
\includegraphics[height=6.5cm, width=10cm]{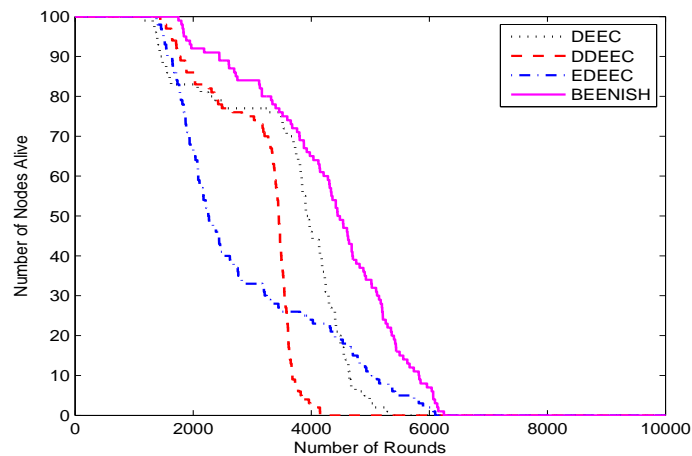}
\caption{Alive Nodes During Network Lifetime}
\end{figure}

\begin{figure}[!h]
\centering
\includegraphics[height=6.5cm, width=10cm]{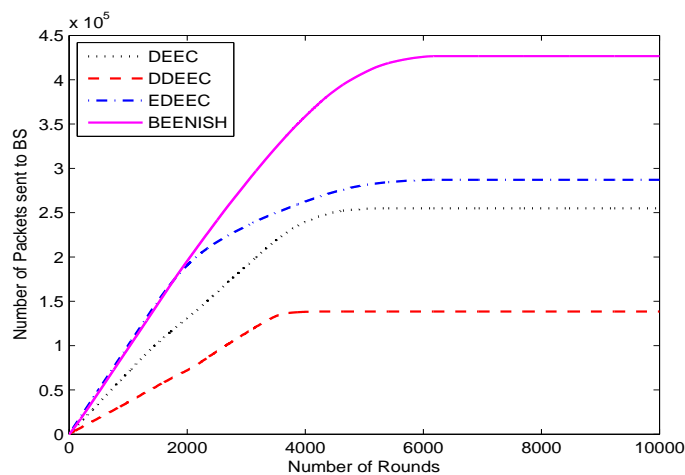}
\caption{Packets sent to BS}
\end{figure}

\section{Conclusion}
Our proposed BEENISH is energy-aware clustering protocol for heterogenous WSNs, with the concept of four types of nodes. Election of CH based on residual and average energy of the network. So, nodes with high energy have more chances to get selected as CH, as compare to the low energy nodes. BEENISH is proved to be the most efficient protocols as compared to DEED, DDEEC and EDEEC for all types of WSNs in terms of stability period, network lifetime and throughput.

\bibliographystyle{plain}

\end{document}